\documentclass[prd,amsmath,amssymb,preprintnumbers,showpacs,showkeys]{revtex4}

\usepackage{amsmath,amssymb,mathrsfs}
\usepackage[dvips]{graphics}
\usepackage{pstcol,pst-3d}

\renewcommand{\square}{\vrule height 1.5ex width 1.2ex depth -.1ex }

\newcommand{\II}{\leavevmode\hbox{\rm{\small1\kern-3.8pt\normalsize1}}}

\newcommand{\RR}{{\mathbb R}}

\newtheorem{Thm}{Theorem}[section]

\numberwithin{equation}{section}

\newcommand{\HH}{{\mathscr H}}

\newcommand{\OO}{{\cal O}}

\newcommand{\kb}{{\boldsymbol{k}}}
\newcommand{\lb}{{\boldsymbol{\ell}}}
\newcommand{\gb}{{\boldsymbol{g}}}

\renewcommand{\Re}{{\rm Re}\,}

\newcommand{\ip}[2]{{\langle #1\mid #2\rangle}}
\newcommand{\ket}[1]{{\vert #1\rangle}}

\newcommand{\Ob}{{\boldsymbol{0}}}

\begin{document}
\title{Null energy conditions in quantum field theory}
\author{Christopher J. Fewster}
\email{cjf3@york.ac.uk}
\author{Thomas A. Roman\footnote{Permanent address: Department
of Physics and Earth Sciences, Central Connecticut State University, New
Britain, CT 06050, U.S.A.}}
\email{roman@ccsu.edu}
\affiliation{Department of Mathematics, University of York, Heslington, York, YO10 5DD, UK}
\date{September 10, 2002. Revised November 25, 2002}
\begin{abstract} 
For the quantised, massless, minimally coupled real scalar field in
four-dimensional Minkowski space, we show (by an explicit construction) 
that weighted averages
of the null-contracted stress-energy tensor along null geodesics
are unbounded from below on the class of Hadamard states.
Thus there are no quantum inequalities along null geodesics in
four-dimensional Minkowski spacetime.
This is in contrast to the case for two-dimensional flat
spacetime, where such inequalities do exist. We discuss in detail the
properties of the quantum states used in our analysis, and also show
that the renormalized expectation value of the stress energy tensor
evaluated in these states satisfies the averaged null energy condition
(as expected), despite the nonexistence of a null-averaged
quantum inequality. However, we also show that in any globally hyperbolic 
spacetime the null-contracted
stress energy averaged over a {\it timelike} worldline {\it does} satisfy
a quantum inequality bound (for both massive and massless fields). 
We comment briefly on the implications of
our results for singularity theorems.
\end{abstract}
\preprint{ESI 1205 (2002)}
\maketitle

\section{Introduction}

The field equations of general relativity have little or no predictive
power in the absence of some notion of what metrics or stress-energy
tensors are to be regarded as physically reasonable. In classical
general relativity it has proved profitable to require the
stress-energy tensor to satisfy one or more of the so-called energy
conditions: in particular, the Hawking--Penrose singularity theorems~\cite{P,HawkPen70}
and the positive mass theorem~\cite{SchYau81,SchYau82} are proved under such
assumptions.

The present paper addresses the status of the null energy condition
(NEC) in quantum field theory. The classical NEC is the requirement
that the stress-energy tensor $T_{ab}$ should obey
$T_{ab}\ell^a\ell^b\ge 0$ for all null vectors $\ell^a$ and at every
spacetime point. Although this condition is satisfied by many
classical matter models, including the minimally coupled scalar field
and the electromagnetic field~\footnote{A matter model violating the
NEC is the nonminimally coupled scalar field~\cite{BarVis}.},  it is
known, however, that this condition is violated by quantum fields. In
fact, the expectation $\langle T^{\rm ren}_{ab}\ell^a\ell^b\rangle_\omega$ of the renormalised
null-contracted stress-energy is unbounded from below as a function of
the quantum state $\omega$. Exactly the same phenomenon afflicts the
weak energy condition (WEC) which, classically, requires that
$T_{ab}v^a v^b\ge 0$ for all timelike vectors $v^a$. In this case, it
is known that the renormalised energy density is still subject to
constraints on its averages along timelike curves. For example, the
massless real scalar field in $n$-dimensional Minkowski space obeys~\cite{Flan2d,FewsterEveson}
\begin{equation}
\int dt\, \langle :T_{00}:\rangle_{\omega}(t,\Ob)
g(t)^2 \ge -c_n\int du\, u^n |\widehat{g}(u)|^2
\end{equation}
for all Hadamard states~\footnote{In Minkowski space, a state is
Hadamard if and only if its normal ordered two-point function $\langle
:\Phi(x)\Phi(x'):\rangle_\omega$ is smooth in $x$ and $x'$.}  $\omega$
and any smooth, real-valued compactly supported function $g$,
where $\widehat{g}$ is the Fourier transform of $g$ (see Eq.~(\ref{eq:FTconv})),
and the $c_n$ are explicitly known constants independent of $\omega$
and $g$. Such constraints are known as {\em quantum
weak energy inequalities} (QWEIs) and appear to be the vestiges of the
weak energy condition in quantum field theory~\footnote{It would be
interesting to understand whether, conversely, the weak energy
condition can be regarded as a classical limit of the QWEIs.}.  Over
the past decade, QWEIs have been developed in a variety of
circumstances
\cite{Flan2d,FewsterEveson,FR95,FR97,PFGQI,MP-EM,FewsterTeo1,Vollick,Flan2d2}
and are known to
hold for the minimally coupled scalar field, the Dirac field and the
electromagnetic and Proca fields in great generality \cite{AGWQI,FV,FewPfen}.
The QWEIs also imply that the averaged weak energy condition (AWEC)
\begin{equation}
\int dt\, \langle :T_{00}:\rangle_{\omega}(t,\Ob) \ge 0
\end{equation}
holds at least for Hadamard states $\omega$ for which the integral converges
absolutely~\footnote{For general Hadamard states one has the weaker result
${\rm lim\,inf}\,_{t_0 \to+\infty} \int dt\, \langle :T_{00}:\rangle_{\omega}(t,\boldsymbol{0})
g(t/t_0)^2 \ge 0$.}.

It is natural to enquire whether similar vestiges of the NEC persist in
quantum field theory.
This is particularly relevant to attempts to generalise the singularity
theorems to quantised matter fields as it is the NEC which 
is assumed in the Penrose theorem \cite{P}.  
While the final stages of gravitational collapse presumably require a
full theory of quantum gravitation for their description, the early
stages can certainly be treated within quantum field theory on a fixed
curved spacetime. The question to be addressed is whether an initially
contracting matter distribution will continue to do so and it is here
that the NEC (or its variants) appears in the classical arguments.
Accordingly, it is important to understand whether the NEC has an
analogue in quantum field theory.
For massless fields in two-dimensional Minkowski space, this question was
answered affirmatively in Ref.~\cite{FR95},
using a Lorentzian sampling function. The bound has the form
\begin{equation}
\frac{\lambda_0}{\pi} \int_{-\infty}^{\infty}d{\lambda}\,
\frac{\langle :T_{ab}:l^a l^b \rangle_\omega(\gamma(\lambda))}{{\lambda}^2+{{\lambda}_0}^2}
\geq -\frac{1}{16\pi {{\lambda_0}^2}}\, ,
                                                  \label{eq:NQI1}
\end{equation}
for all $\lambda_0>0$ and a large class of states~\footnote{Although this class
was not precisely delineated in Ref.~\cite{FR95}, one expects the bound to hold
for all
Hadamard states for which the left-hand side exists.} $\omega$, where
$\gamma$ is
an affinely parametrized null geodesic with tangent vector
$\ell^a=(d\gamma/d\lambda)^a$.
It can be easily seen that this bound is invariant
under a rescaling of the affine parameter.
If we now take the limit of Eq.~(\ref{eq:NQI1}) as
${\lambda_0}\rightarrow\infty$, which corresponds to sampling the entire
null geodesic, we get the ANEC~\footnote{At least for those states in which the
integral in Eq.~(\ref{eq:ANEC2D}) converges absolutely.}:
\begin{equation}
\int_{-\infty}^{\infty}d{\lambda}\,
\langle :T_{ab}:l^a l^b \rangle_\omega  \geq 0  \label{eq:ANEC2D}\,.
\end{equation}

Reference~\cite{FR95} left open the question of whether an analogous
QNEI exists in spacetime dimensions other than two.  The techniques
used there to obtain a timelike worldline QI in four dimensions could
not be employed to derive a similar QNEI, starting with null geodesics
{\it ab initio}, because the former derivation was based upon a mode
expansion in the timelike observer's rest frame. There are also technical
problems which obstruct the adaptation of the arguments of Ref.~\cite{AGWQI} to null
worldlines (see the remark following Theorem~\ref{thm:nontriv}). 
In addition, Ref.~\cite{FR95} noted a
potential problem: any such inequality involving an average
along a null geodesic would have to be invariant under rescaling of
the affine parameter (amounting to the replacements $\lambda\mapsto
\lambda/\sigma$, $\lambda_0\mapsto \lambda_0/\sigma$ and
$\ell^a\mapsto \sigma \ell^a$ in Eq.~(\ref{eq:NQI1})) to be physically
meaningful. While the left-hand side of Eq.~(\ref{eq:NQI1}) scales as
$\sigma^2$, one might expect (on dimensional grounds) that the
right-hand side of such a bound would behave like $\lambda_0^{-d}$,
where $d$ is the spacetime dimension, and therefore scale as
$\sigma^d$. This hints that the extension of QNEIs to
spacetime dimensions $d>2$ might be problematic. 
(Of course, these arguments would not
apply in the presence of a mass or some other geometrical length scale
--- see Ref.~\cite{FR96_EBH}.)

In this paper, we consider worldline averages of the null-contracted
stress-energy tensor of the form
\begin{equation}
\langle \rho (f)\rangle_{\omega}
=\int d\lambda\, \langle :T_{ab}:l^a l^b \rangle_\omega(\gamma(\lambda))
\end{equation}
where $\gamma(\lambda)$ is a smooth causal curve and
$\ell^a$ is a smooth null vector field defined on $\gamma$. 
First, in Sect.~\ref{sect:nogo} we study the case in which $\gamma$
is an affinely parametrised null geodesic in four-dimensional Minkowski space
with tangent vector $\ell^a=(d\gamma/d\lambda)^a$. By an explicit construction,
we show that $\langle \rho (f)\rangle_{\omega}$ is
unbounded from below as $\omega$ varies among the class of Hadamard states
of the massless minimally coupled scalar field. Thus there are no
null-worldline QNEIs in four-dimensional Minkowski space. Although we
consider only the massless field, we comment that our results generalise
directly to the massive case. Our construction involves a sequence of
states, each of which is a superposition of the vacuum with a multimode
two-particle state. A closely related construction has recently been used
in Ref.~\cite{FHR} to prove the nonexistence of spatially averaged quantum
inequalities in four-dimensional Minkowski space.

It would be incorrect, however, to conclude from the above result that 
the null-contracted stress-energy tensor is completely unconstrained
in quantum field theory. In Sect.~\ref{sect:nontriv} we consider the
averages $\langle \rho (f)\rangle_{\omega}$ for smooth
timelike $\gamma$ in an arbitrary globally hyperbolic spacetime and
for any smooth null vector field $\ell^a$. For both massive and massless fields,
these quantities do obey lower bounds ---
which we call timelike worldline QNEIs --- as a
direct consequence of the arguments used in Ref.~\cite{AGWQI}. We
evaluate our bound explicitly for the case of four-dimensional
Minkowski space. Taken together with the results of Sec.~\ref{sect:nogo},
we see that large negative values of the null-contracted stress-energy
tensor on one null geodesic must be compensated by positive values on
neighbouring geodesics, because the transverse extent of the negative
values is constrained by timelike worldline QNEIs. In the conclusion, 
we briefly speculate on the significance of these results for attempts
to derive singularity theorems for quantised matter.


\section{Nonexistence of null-worldline QNEIs}
\label{sect:nogo}

\subsection{Nonexistence result}

We consider a massless minimally coupled real scalar field in
1+3-dimensional Minkowski
space, with signature ${+}{-}{-}{-}$. We employ units with $\hbar=c=1$.
The quantum field is given by
\begin{equation}
\Phi(x) = \int\frac{d^3\kb}{(2\pi)^3(2\omega)^{1/2}}
\left(a(\kb)e^{-ik_a x^a} + a^\dagger(\kb) e^{ik_a x^a}\right)\,,
\end{equation}
in which $k^a=(\omega,\kb)$ with $\omega=\|\kb\|$, the magnitude of $\kb$. The
canonical commutation relations are
\begin{eqnarray}
[a(\kb),a(\kb')] &=& 0 \nonumber\\
{}[a(\kb),a^\dagger(\kb')] &=& (2\pi)^3 \delta(\kb-\kb')\,,
\end{eqnarray}
and our convention for Fourier transformation is
\begin{equation}
\widehat{f}(u)= \int dt\,e^{-iut} f(t)\,.
\label{eq:FTconv}
\end{equation}

Now let $f$ be any smooth nonnegative function of compact support, normalised
so that
\begin{equation}
\int d\lambda f(\lambda) =1\,,
\end{equation}
and, for some fixed future-pointing null vector $\ell^a$, let 
$\gamma(\lambda)$ be the null geodesic $\gamma(\lambda)^a =
\lambda\ell^a$. For simplicity, we
will assume that the three-vector part of $\ell^a$ has unit length (in our frame of
reference), so $\ell^a=(1,\lb)$ with $\|\lb\|=1$. We will
consider the averaged quantity
\begin{equation}
\langle\rho(f)\rangle_\omega = \int d\lambda f(\lambda) \langle
:T_{ab}:\ell^a\ell^b\rangle_\omega(\gamma(\lambda))\,,
\label{eq:rhofom}
\end{equation}
which corresponds to a weighted average of the null-contracted stress
energy tensor along $\gamma$. If $\omega$ is a Hadamard state, the
renormalised contracted stress
tensor is a smooth function on spacetime, so the above integral will
certainly converge. In order to establish a quantum null energy
inequality, one would need to bound $\langle\rho(f)\rangle_\omega$ from below as
$\omega$ ranges over the class of Hadamard states; however,
this is not possible, as we now show.

\begin{Thm} \label{thm:main}
The quantity $\langle\rho(f)\rangle_\omega$ is unbounded from below as
$\omega$ varies over the class of Hadamard states.
\end{Thm}
{\em Proof:} We will construct a family of vector Hadamard states
$\omega_\alpha$
($\alpha\in(0,1)$) with the property that
$\langle\rho(f)\rangle_{\omega_\alpha}\to -\infty$ as $\alpha\to 0$.
We begin by choosing a fixed $\Lambda_0>0$ such that
$\Re \widehat{f}$ is nonnegative on the interval
$[-2\Lambda_0,2\Lambda_0]$. To see that this is possible, we observe
that
\begin{equation}
\widehat{f}(0) = \int d\lambda f(\lambda) =1\,,
\end{equation}
which, by continuity, implies that $\Re\widehat{f}$ is positive in some
neighbourhood of the origin.

Next, let $\sigma$ and $\nu$ be fixed positive numbers with
$2\nu+3/2<\sigma<2\nu+2$.
For each $\alpha\in(0,1)$, we define a
`vacuum-plus-two-particle' vector
\begin{equation}
\psi_\alpha = N_\alpha\left[
\ket{0} + \int \frac{d^3\kb}{(2\pi)^3}\,\frac{d^3\kb'}{(2\pi)^3}
b_\alpha(\kb,\kb')\ket{\kb,\kb'}\right]\,,
\end{equation}
where $N_\alpha$ is a normalisation constant ensuring $\|\psi_\alpha\|=1$
and
\begin{equation}
b_\alpha(\kb,\kb') =
\alpha^\sigma\vartheta(\Lambda-k)\vartheta(\Lambda-k')\chi_\alpha(\theta)
\chi_\alpha(\theta')
B(k_a\ell^a,k'_a \ell^a)(kk')^{\nu-1/2}\,.
\label{eq:balpha_def}
\end{equation}
Here, $\vartheta$ is the usual Heaviside step function,
$\Lambda=\Lambda_0/\alpha$ will be called the momentum cut-off and
\begin{equation}
\chi_\alpha(\theta) = \left\{\begin{array}{cl} 1\,, & \cos\theta>1-\alpha\\
0\,, & {\rm otherwise,}\end{array}\right.
\end{equation}
where $\theta$ (respectively, $\theta'$) is the angle between $\kb$ (resp.,
$\kb'$) and $\lb$. We choose $B:\RR^+\times\RR^+\to\RR$ to be
(a)~symmetric (i.e., $B(u,u')=B(u',u)$); (b)~jointly continuous in $u$ and
$u'$; (c)~everywhere nonnegative~\footnote{Although $B$ is real-valued,
we shall write complex conjugations where they would be appropriate
for complex $B$.} and strictly positive near $u=u'=0$; and
(d)~normalised so that
\begin{equation}
\Lambda_0^{4\nu+2}\int_0^\infty du \int_0^\infty du' \,|B(u,u')|^2 =1\,.
\label{eq:Bnorm}
\end{equation}
(The prefactor ensures dimensional consistency.)
An example of a function meeting these requirements is
$B(u,u') = \Lambda_0^{-(2\nu+1)}e^{-(u+u')/2}$. 
We wish to
emphasise, however, that there are many functions (and hence many vectors
$\psi_\alpha$) with the properties we require. We will use $\omega_\alpha$
to denote the
state induced by $\psi_\alpha$ so that
$\langle A \rangle_{\omega_\alpha} = \ip{\psi_\alpha}{A\psi_\alpha}$.

\begin{figure}
\begin{center}
\resizebox*{2.5 in}{!}{\includegraphics{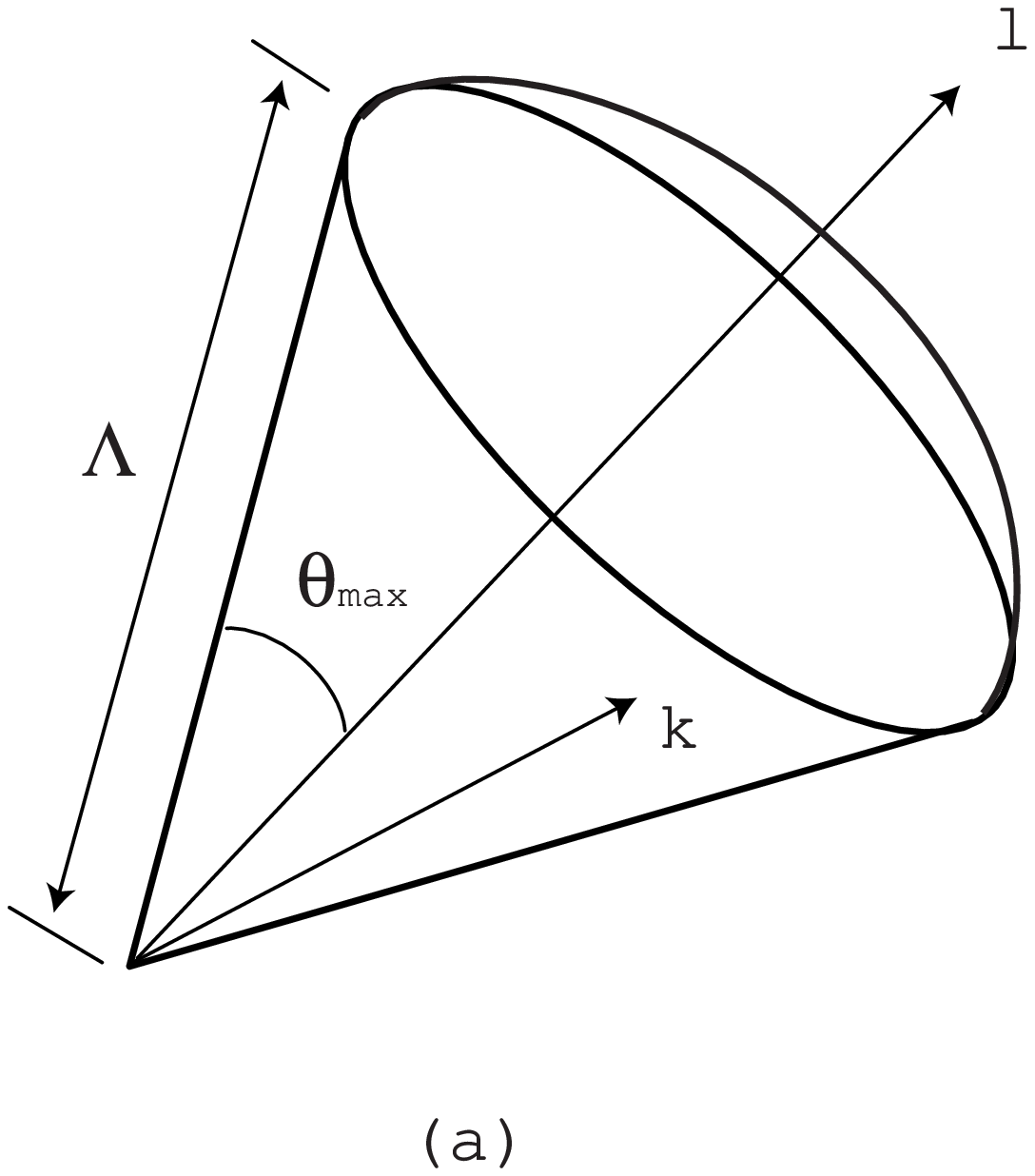}}
\hspace{0.5in}\resizebox*{2.5 in}{!}{\includegraphics{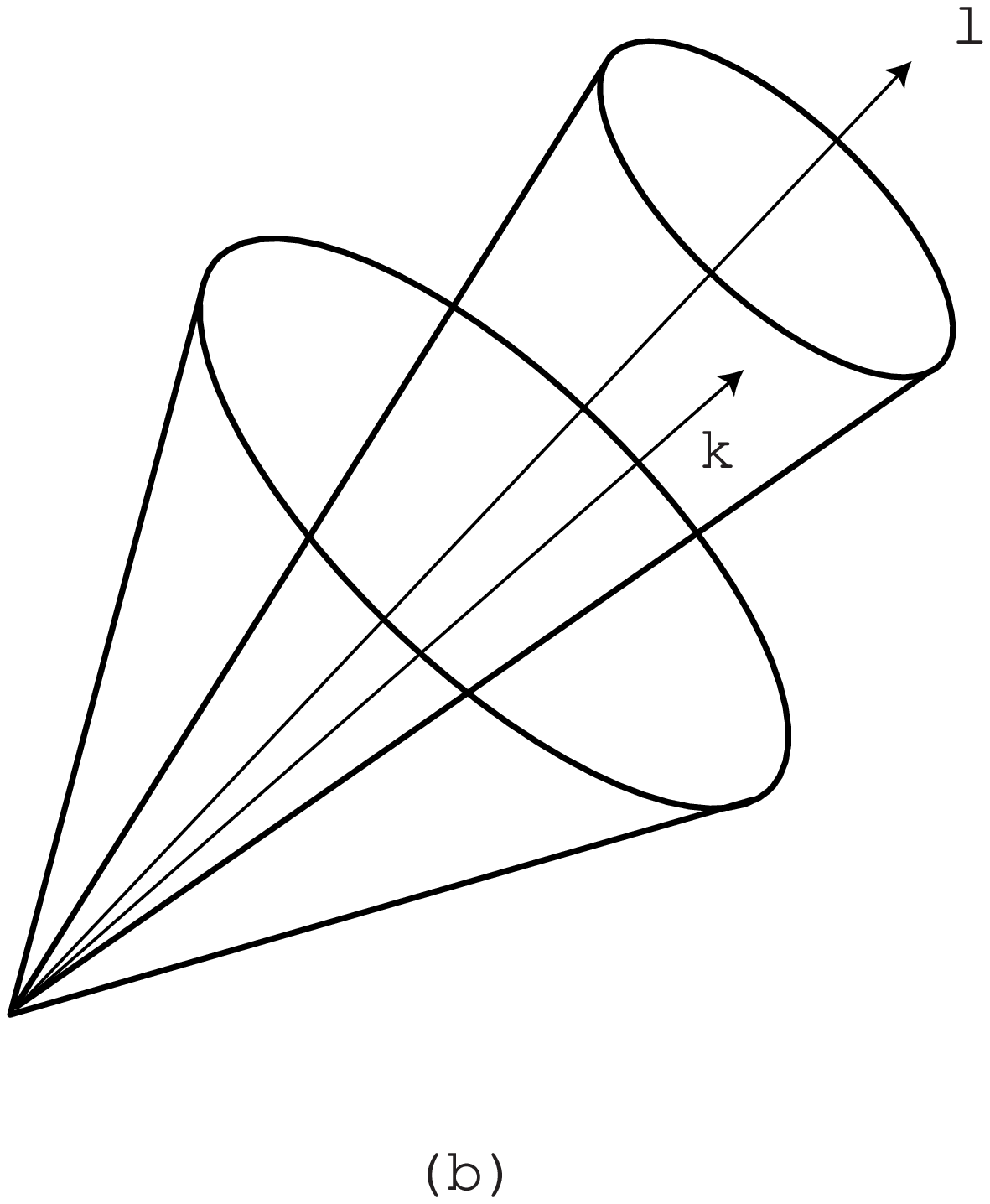}}
\end{center}
\caption{(a) Only modes lying ``inside the cone'', [i.e.,
those whose three-momenta make an angle less than 
$\theta_{\rm max}= \cos^{-1} \, (1- \alpha)$ with $\lb$]
are excited; (b) The cones lengthen and tighten around $\lb$
as $\alpha \rightarrow 0$.}
\label{fig:cones}
\end{figure}

Let us note various features of this family of states. First, the momentum
cut-off
ensures that no modes of momentum greater than
$\Lambda=\Lambda_0/\alpha$ are excited. Second, the effect of the
$\chi_\alpha$ factors is to ensure that modes can only be excited if
their three-momenta make an angle less than $\cos^{-1}(1-\alpha)$ with the
direction $\lb$. The excited mode three-momenta therefore lie in the solid
sector formed by the intersection of a ball of radius $\Lambda_0/\alpha$
(centre the origin) with a cone of opening angle $\cos^{-1}(1-\alpha)$ about
$\lb$ (with apex
at the origin). As $\alpha\to 0$, this solid sector lengthens and
tightens up along the direction $\lb$, so the four-momenta of excited modes
become more
and more parallel to $\ell^a$, the tangent vector to the null line along
which we are averaging. See Fig.~\ref{fig:cones}.

The third feature of interest concerns the amplitude of the two-particle
contribution. Choosing the normalisation constant to be
\begin{equation}
N_\alpha = \left[1+2\int\frac{d^3\kb}{(2\pi)^3}\,\frac{d^3\kb'}{(2\pi)^3}
|b_\alpha(\kb,\kb')|^2\right]^{-1/2}\,,
\label{eq:normconst}
\end{equation}
we note that, for a null vector $k^a=(k,\kb)$, the quantity $\ell^a
k_a$ appearing in Eq.~(\ref{eq:balpha_def}) is equal to 
$k(1-\cos\theta)$, where $\theta$ is the angle between
$\lb$ and $\kb$. We therefore perform the $\kb$ and $\kb'$ integrals
in Eq.~(\ref{eq:normconst}) by 
adopting spherical polar coordinates about $\lb$,
integrating out the trivial azimuthal dependence and then changing
variables to $\beta=1-\cos\theta$, $\beta'=1-\cos\theta'$. This yields
\begin{eqnarray}
\int \frac{d^3\kb}{(2\pi)^3}\,\frac{d^3\kb'}{(2\pi)^3}
|b_\alpha(\kb,\kb')|^2 &=&
\frac{\alpha^{2\sigma}}{(2\pi)^4}\int_0^\Lambda dk\int_0^\Lambda dk'
(kk')^{2\nu+1}\int_0^\alpha d\beta \int_0^\alpha d\beta'
\,|B(k\beta,k'\beta')|^2\nonumber\\
&=& \frac{\alpha^{2\sigma-2-4\nu}}{(2\pi)^{4}}\int_0^{\Lambda_0}dv
\int_0^{\Lambda_0}dv' (vv')^{2\nu}\int_0^v du\int_0^{v'} du'
|B(u,u')|^2\nonumber\\
&\le& \frac{\alpha^{2(\sigma-2\nu-1)}}{(2\pi)^{4}(2\nu+1)^2}\,,
\label{eq:intbsq}
\end{eqnarray}
where we have made the further changes of variable
$u=k\beta$, $u'=k'\beta'$, $v=k\alpha$, $v'=k'\alpha$ and
used the normalisation property Eq.~(\ref{eq:Bnorm}) of $B$.
Because $\sigma>2\nu+3/2>2\nu+1$, we see that the right-hand side
of Eq.~(\ref{eq:intbsq}) tends to zero as $\alpha\to 0$. By
Eq.~(\ref{eq:normconst}) we now have $N_\alpha\to 1$ as $\alpha\to 0$;
since the left-hand side of Eq.~(\ref{eq:intbsq}) is equal to
$\|N_\alpha^{-1}\psi_\alpha-\ket{0}\|^2$,  we also see that
the states $\psi_\alpha$ are in fact converging to the vacuum
vector $\ket{0}$. As we shall see, this does not entail that the
normal-ordered energy density is converging to zero. (See also the
discussion in Sect.~\ref{sect:cvgnce}.)

The remaining properties of our family of states concern the corresponding 
normal ordered two-point functions, given by
\begin{eqnarray}
\langle :\Phi(x) \Phi(x'):\rangle_{\omega_\alpha}
&=& 2N_\alpha^2\,\Re
\int\frac{d^3\kb}{(2\pi)^3}\,\frac{d^3\kb'}{(2\pi)^3}
\frac{1}{\sqrt{\omega\omega'}}
\left[c_\alpha(\kb,\kb') e^{i(x^a k_a-{x'}^a k'_a)} \right.\nonumber\\
&&\qquad\qquad\qquad\qquad \left.+ b_\alpha(\kb,\kb')e^{-i(x^a k_a+{x'}^a k'_a)}
\right]\,,
\end{eqnarray}
where
\begin{equation}
c_\alpha(\kb,\kb') = 2\int \frac{d^3\kb_1}{(2\pi)^3}
\overline{b_\alpha(\kb_1,\kb)}b_\alpha(\kb_1,\kb')\,.
\end{equation}
Using the same changes of variable as above, we find
\begin{equation}
c_\alpha(\kb,\kb') =
\alpha^{2\sigma-(2\nu+1)}\vartheta(\Lambda-k)\vartheta(\Lambda-k')
\chi_\alpha(\theta)\chi_\alpha(\theta')
C(k_a\ell^a,k'_a \ell^a)(kk')^{\nu-1/2}\,,
\end{equation}
where $C(u,u')=C(u',u)$ is given by
\begin{eqnarray}
C(u,u')&=& 2\alpha^{2\nu+1}\int\frac{d^3\kb_1}{(2\pi)^3}
\vartheta(\Lambda-k_1)\chi_\alpha(\theta_1)
(k_1^2)^{\nu-1/2}\overline{B(k^a_1\ell_a,u)}B(k^a_1\ell_a,u') \nonumber\\
&=& \frac{\alpha^{2\nu+1}}{2\pi^2}\int_0^{\Lambda}
dk_1\,k_1^{2\nu+1}\int_0^\alpha
d\beta\,\overline{B(k_1\beta_1,u)}B(k_1\beta_1,u')\nonumber\\
&=& \frac{1}{2\pi^2}\int_0^{\Lambda_0}
dv\,v^{2\nu}\int_0^v du_1 \overline{B(u_1,u)}B(u_1,u')\,.
\end{eqnarray}
Rearranging the order of integration, this becomes
\begin{equation}
C(u,u') = \frac{1}{2\pi^2(2\nu+1)}\int_0^{\Lambda_0} du_1\,
u_1^{2\nu+1}\overline{B(u_1,u)}B(u_1,u')
\end{equation}
and we may conclude that (i) $C$ is jointly continuous in $u$ and $u'$ by
joint continuity of $B$ and compactness of $[0,\Lambda_0]$; (ii)
$C$ has the same engineering dimension as $B$; (iii) $C(u,u')\ge 0$ for all $u,u'$
and, crucially, (iv) that
the exponent of $\alpha$ in $c_\alpha(\kb,\kb')$ differs from that in the
corresponding expression for $b_\alpha(\kb,\kb')$. Furthermore, since both
$b_\alpha$ and $c_\alpha$ have momentum cut-offs, it is evident that the
normal ordered two-point function is smooth (because one may differentiate under the
integral sign as often as required
to obtain finite derivatives). Accordingly, each $\omega_\alpha$ is a
Hadamard state.

The null-contracted normal ordered energy density $\langle\rho(f)\rangle_{\omega_\alpha}$ 
is obtained by differentiating the normal ordered
two-point function
\begin{eqnarray}
\langle :T_{ab}(x):\ell^a\ell^b\rangle_{\omega_\alpha} &=&
\langle :\ell^a\nabla_a \Phi(x) \ell^b
\nabla_b\Phi(x):\rangle_{\omega_\alpha} \nonumber \\
&=& 2N_\alpha^2\,\Re
\int\frac{d^3\kb}{(2\pi)^3}\,\frac{d^3\kb'}{(2\pi)^3}
\frac{\ell^a k_a \ell^{b} k'_b}{\sqrt{\omega\omega'}}
\left[c_\alpha(\kb,\kb') e^{ix^a (k_a- k'_a)} \right. \nonumber\\
&&\qquad\qquad\qquad\qquad\left.
- b_\alpha(\kb,\kb')e^{-ix^a
(k_a+ k'_a)}
\right]\,,
\end{eqnarray}
and substituting into Eq.~(\ref{eq:rhofom}). Noting, for any $K_a$, that
\begin{equation}
\int d\lambda\, f(\lambda) e^{-i\gamma(\lambda)^a K_a} =\widehat{f}(\ell^a K_a)
\,,
\end{equation}
we may write $\langle \rho(f)\rangle_{\omega_\alpha}=\rho_1(f)+\rho_2(f)$, where
\begin{equation}
\rho_1(f) = 2N_\alpha^2\,\Re
\int\frac{d^3\kb}{(2\pi)^3}\,\frac{d^3\kb'}{(2\pi)^3}
\frac{\ell^a k_a \ell^b k'_b}{\sqrt{\omega\omega'}}
\widehat{f}(\ell^a k'_a -\ell^a k_a) c_\alpha(\kb,\kb')
\end{equation}
and
\begin{equation}
\rho_2(f) = -2N_\alpha^2 \,\Re
\int\frac{d^3\kb}{(2\pi)^3}\,\frac{d^3\kb'}{(2\pi)^3}
\frac{\ell^a k_a \ell^b k'_b}{\sqrt{\omega\omega'}}
\widehat{f}(\ell^a k_a +\ell^a k'_a) b_\alpha(\kb,\kb')\,.
\end{equation}
Our aim is now to show that $\rho_1(f)\to 0$ and $\rho_2(f)\to-\infty$
in the limit $\alpha\to 0$. Taking the dominant contribution $\rho_2(f)$ first,
and making the same changes of variable as before we may calculate
\begin{eqnarray}
\rho_2(f) &=& -\frac{N_\alpha^2\alpha^\sigma}{(2\pi)^4} \Re
\int_0^\Lambda dk\int_0^\Lambda dk'
(kk')^{\nu+2}\int_0^\alpha d\beta \int_0^\alpha d\beta'\,\beta\beta'
B(k\beta,k'\beta')\widehat{f}(k\beta+k'\beta')   \nonumber\\
&=& -\frac{N_\alpha^2\alpha^\sigma}{(2\pi)^4}
\int_0^\Lambda dk\int_0^\Lambda dk' (kk')^\nu \varphi(k\alpha,k'\alpha)\nonumber\\
&=&-\frac{N_\alpha^2\alpha^{\sigma-2(\nu+1)}}{(2\pi)^4}
\int_0^{\Lambda_0} dv\int_0^{\Lambda_0} dv' (vv')^\nu \varphi(v,v')\,,
\label{eq:rho2final}
\end{eqnarray}
where
\begin{equation}
\varphi(v,v') = \Re \int_0^v du \int_0^{v'} du' uu'B(u,u')
\widehat{f}(u+u')\,.
\label{eq:di0}
\end{equation}
Recalling that $\Re\widehat{f}$ is strictly positive on the interval
$[-2\Lambda_0,2\Lambda_0]$, and that $B(u,u')$ is nonnegative and strictly
positive for small $u,u'$, it follows that
$\varphi(v,v')$ is nonnegative for $v,v'\in[0, \Lambda_0]$ and that
the right-hand side of~(\ref{eq:rho2final}) is strictly negative.
We therefore have $\rho_2(f)\to-\infty$ in the limit $\alpha\to 0$,
because $\sigma<2\nu+2$ and $N_\alpha\to 1$.

Turning to the remaining contribution $\rho_1(f)$, 
we may use a similar analysis to obtain
\begin{equation}
\rho_1(f) = \frac{N_\alpha^2\alpha^{2\sigma-(4\nu+3)}}{(2\pi)^4}
\int_0^{\Lambda_0} dv\int_0^{\Lambda_0} dv'\, (vv')^\nu \psi(v,v')\,,
\label{eq:di1}
\end{equation}
where
\begin{equation}
\psi(v,v') = \Re \int_0^v du \int_0^{v'} du'\, uu'C(u,u')
\widehat{f}(u'-u)\,.
\label{eq:di2}
\end{equation}
Since $C$ is jointly continuous, the double integrals in Eqs.~(\ref{eq:di1}) and~(\ref{eq:di2}) 
exist and are finite; we may now conclude
that $\rho_1(f)\to 0$ in the limit $\alpha\to 0$, because
$\sigma>2\nu+3/2$ and $N_\alpha\to 1$.

Summarising, we have shown that
$\langle\rho(f)\rangle_{\omega_\alpha} \to -\infty$
as $\alpha\to 0$. $\square$

At this point it is worth considering the difference between the
present situation and that studied in Ref.~\cite{FR95}, in which a QNEI
was obtained for massless fields in two-dimensional Minkowski space.
The crucial difference is that, in two-dimensional spacetime,
the one-momenta of any two field modes are either parallel or anti-parallel. The modes which
propagate in the same direction as the chosen null geodesic contribute
nothing to the integral (due to factors of the form $\ell^a k_a$).
The only contribution comes from modes moving
in the opposite direction, and turns out to be bounded. By contrast,
the proof of Theorem~\ref{thm:main} makes essential use of field modes which
are almost, but not exactly, parallel to $\ell^a$.

\subsection{An explicit calculation}

To make the foregoing result more explicit, we consider a simple
example: setting $\nu=1$, we define
\begin{equation}
B(u,u')= \Lambda_0^{-4}
\vartheta(\Lambda_0-u)\vartheta(\Lambda_0-u')\,,
\label{eq:expB}
\end{equation}
which satisfies properties (a), (c) and (d) required in the proof of
Theorem~\ref{thm:main}, but not the joint continuity property
(b). Inspection of the proof reveals, however, that this property was
only used to establish the existence of certain integrals arising in
the derivation, all of which may easily be seen to exist in this case.

\begin{figure}
\begin{center}
\resizebox*{3 in}{!}{\includegraphics*{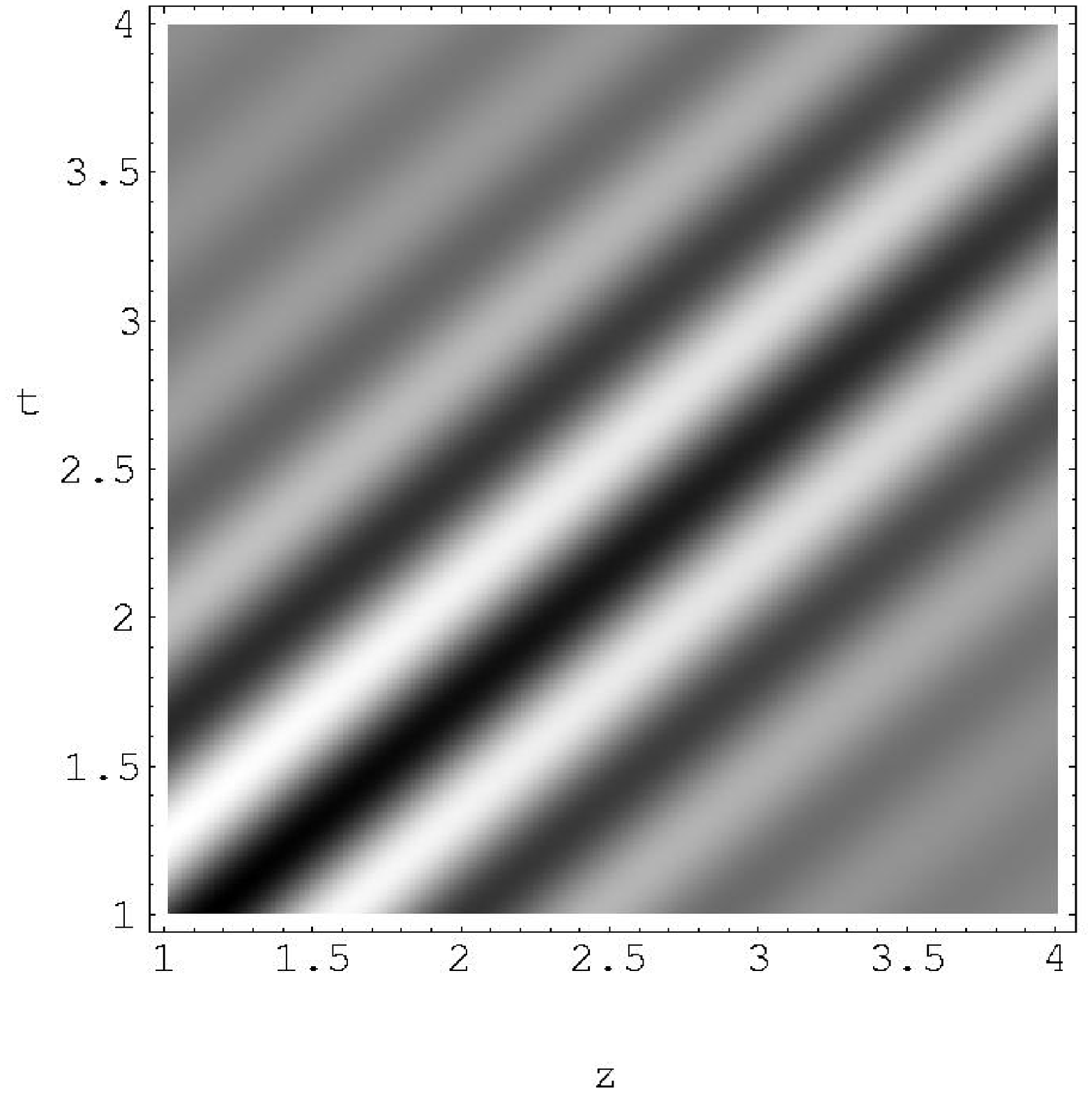}}\hspace{0.2cm}
\resizebox*{3 in}{!}{\includegraphics*{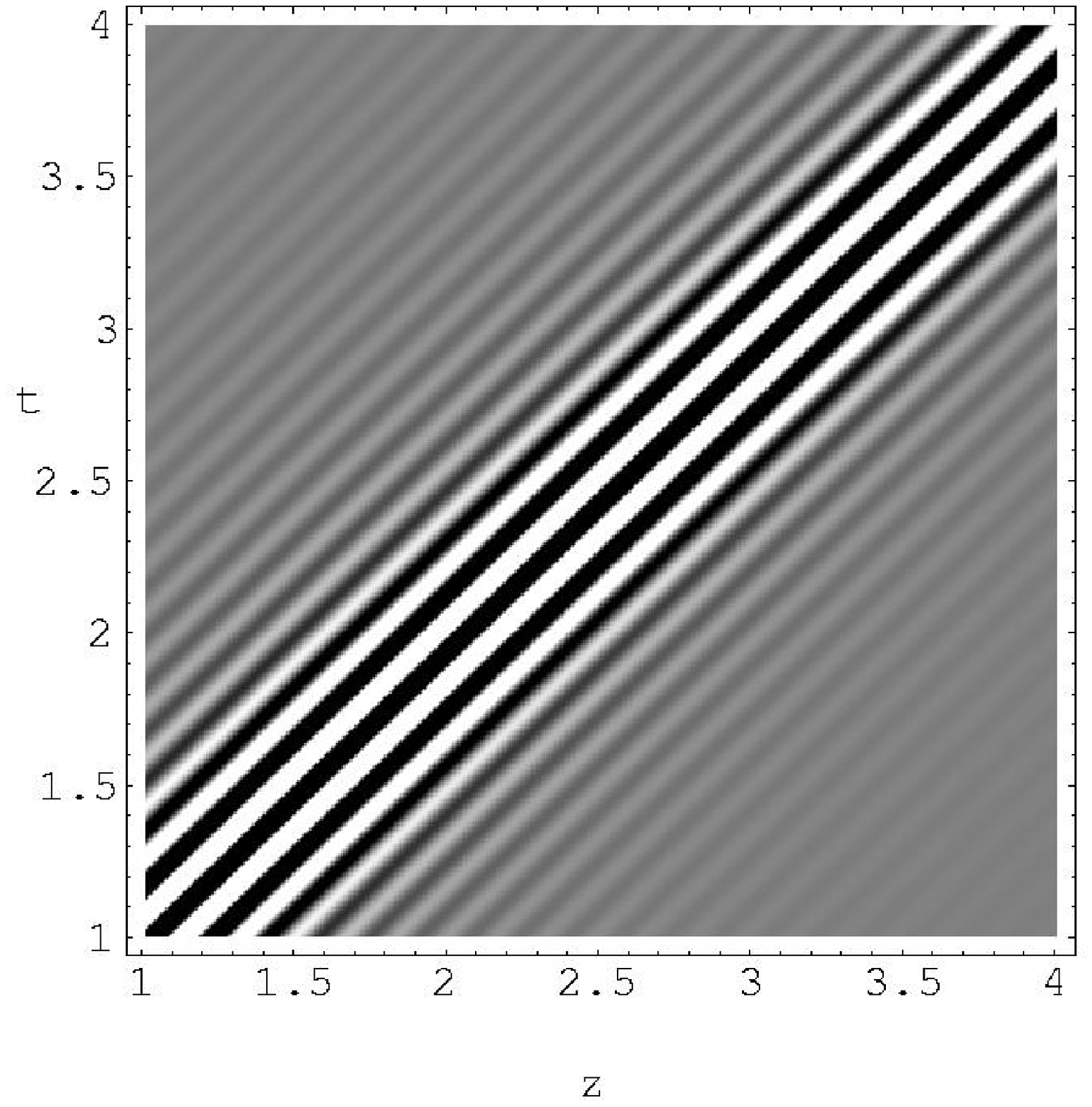}}\\
\resizebox*{3 in}{!}{\includegraphics*{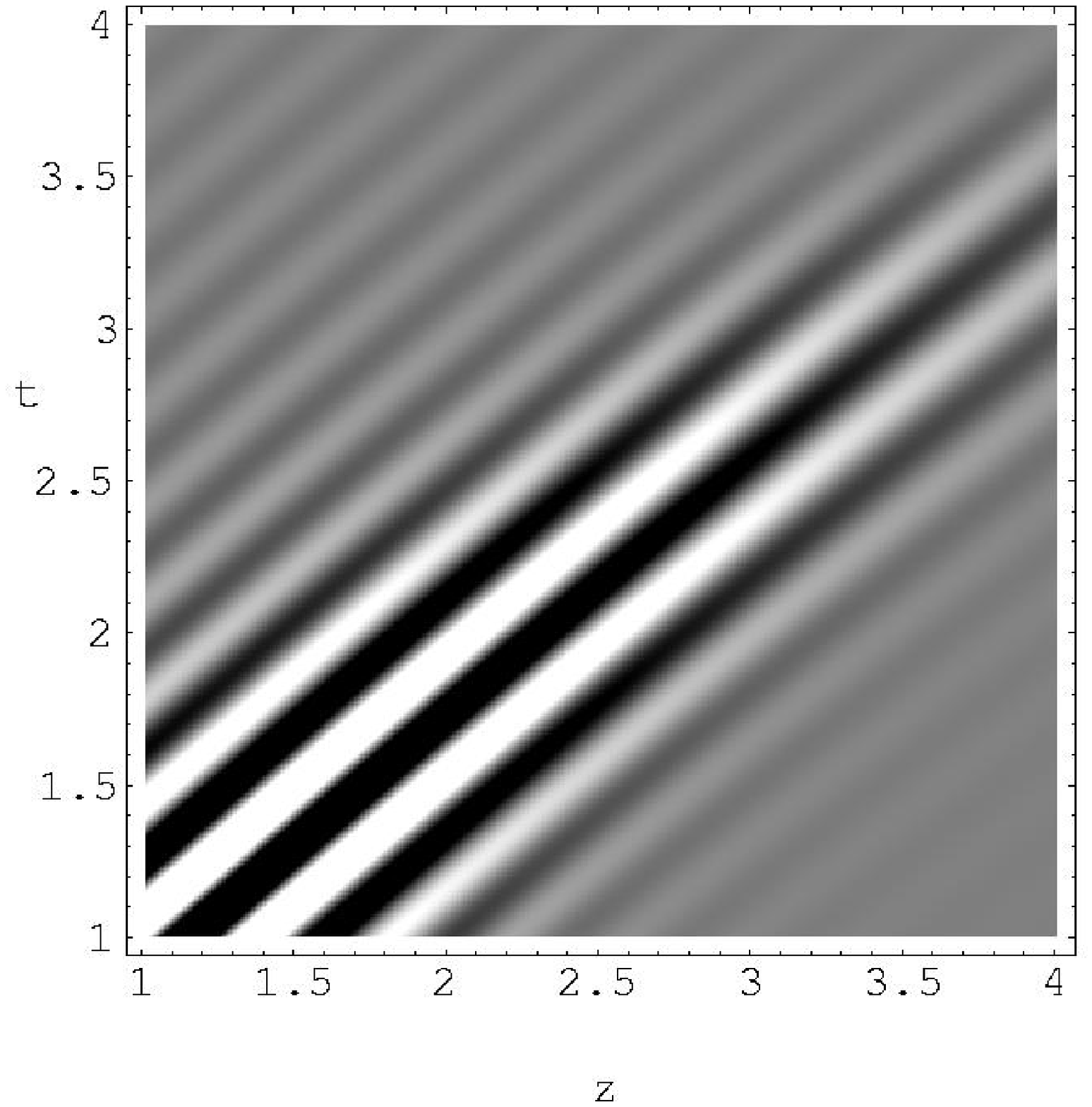}}\hspace{0.2cm}
\resizebox*{3 in}{!}{\includegraphics*{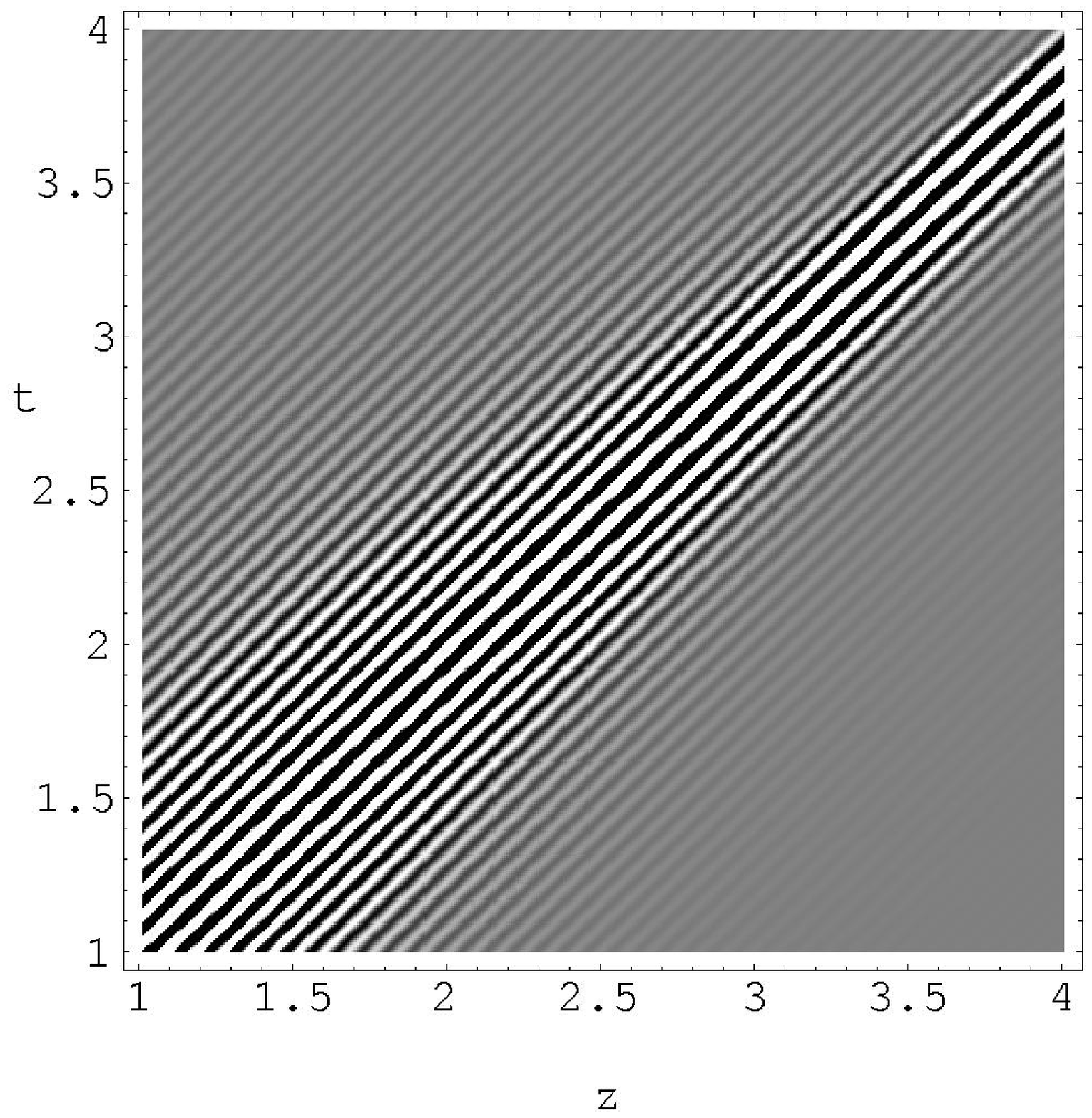}}
\end{center}
\caption{Density plots of $\langle\rho(t,0,0,z)\rangle_{\omega_\alpha}$
for four different parameter choices. The top row corresponds to
$\Lambda_0=1$ while the lower row has
$\Lambda_0=2$; the left-hand column has $\alpha=0.2$, while the right-hand
column has $\alpha=0.05$.
In all four plots, $\sigma=3.75$. Dark and light areas represent negative
and positive values
respectively.}
\label{fig:density_plots}
\end{figure}

The calculations are simplified by the fact that $B(u,u')$ factorises into
functions of $u$ and $u'$.
In particular, one may calculate
\begin{equation}
C(u,u') = \frac{1}{24\pi^2\Lambda_0^4}
\vartheta(\Lambda_0-u)\vartheta(\Lambda_0-u')\,,
\end{equation}
and
\begin{equation}
N_\alpha = \left( 1+ \frac{\alpha^{2\sigma-6}}{128\pi^4}\right)^{-1/2}\,.
\end{equation}
Furthermore, because $B$ is also real-valued,
\begin{equation}
\langle\rho(x)\rangle_{\omega_\alpha} =
\frac{2N_\alpha^2}{\Lambda_0^4} \Re \left(
|F_\alpha(x)|^2\frac{\alpha^{2\sigma-3}}{24\pi^2}-F_\alpha(x)^2\alpha^\sigma
\right)\,,
\label{eq:exprho1}
\end{equation}
where here $\rho(x)$ represents the {\it unsampled} energy density at position $x$,
and 
\begin{equation}
F_\alpha(x) =  \int\frac{d^3\kb}{(2\pi)^3} \,
\frac{\ell^a k_a}{k^{1/2}}\vartheta(\Lambda-k)\chi_\alpha(\theta)
k^{\nu-1/2}
e^{-ik_a x^a}\,.
\end{equation}
In Fig.~\ref{fig:density_plots}, we plot
$\langle\rho(t,0,0,z)\rangle_{\omega_\alpha}$ for the case
$\ell^a=(1,0,0,1)$, in which~(\ref{eq:exprho1}) simplifies to
\begin{equation}
\langle\rho(t,0,0,z)\rangle_{\omega_\alpha} =
\frac{2N_\alpha^2}{16\pi^4\Lambda_0^4} \Re \left(
|\xi_\alpha(t,z)|^2\frac{\alpha^{2\sigma-7}}{24\pi^2}-\xi_\alpha(t,z)^2
\alpha^{\sigma-4}\right)\,,
\end{equation}
where
\begin{equation}
\xi_\alpha(t,z) = \int_0^{\Lambda_0} dv\,
 e^{-iv(t-z)/\alpha}\left[\frac{iv^2}{z} e^{-ivz} +
\frac{v}{z^2}(e^{-ivz}-1)\right]\,.
\end{equation}

These plots share the common feature of an oscillatory fringe
pattern, with dark regions representing negative values for
$\langle\rho(t,0,0,z)\rangle_{\omega_\alpha}$ and light regions
representing positive values. It is no coincidence that these plots
resemble interference patterns: the dominant contribution arises precisely
from interference between the vacuum and two-particle components of
$\psi_\alpha$. The fringes are centred near the null ray parallel to
$\ell^a$ running from the lower left to upper right corners of the
figures, and in fact point along spacelike directions which become
more parallel to $\ell^a$ as $\alpha$ is decreased (moving from the
left-hand to right-hand figure in each row). That these directions 
cannot be timelike follows from the existence of the timelike
worldline QNEIs discussed in Sec.~\ref{sect:nontriv}---an observer
cannot `surf' along a negative energy trough for an indefinite length
of time. See Ref.~\cite{BFR} for similar examples and discussion. The
decreasing fringe separation (as either $\alpha$ decreases or
$\Lambda_0$ increases) indicates a more highly oscilliatory energy
density.

\begin{figure}
\begin{center}
\psset{unit=1in}
\begin{pspicture}(0,0.2)(6.2,3)
\rput[l](0,1.836){\resizebox*{3 in}{!}{\rotatebox{270}{\includegraphics{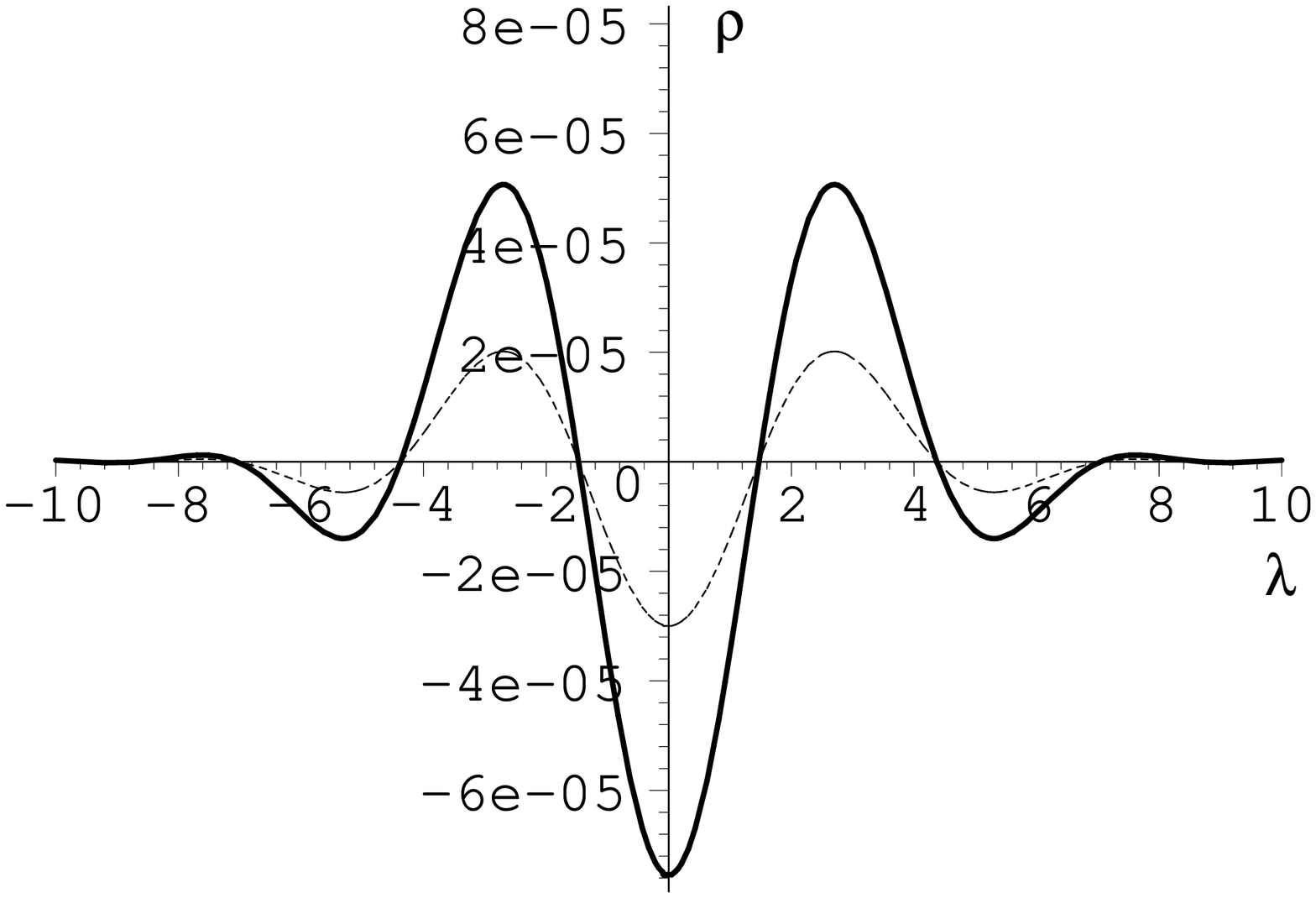}}}}     
\rput[l](3.2,1.8){\resizebox*{3 in}{!}{\rotatebox{270}{\includegraphics{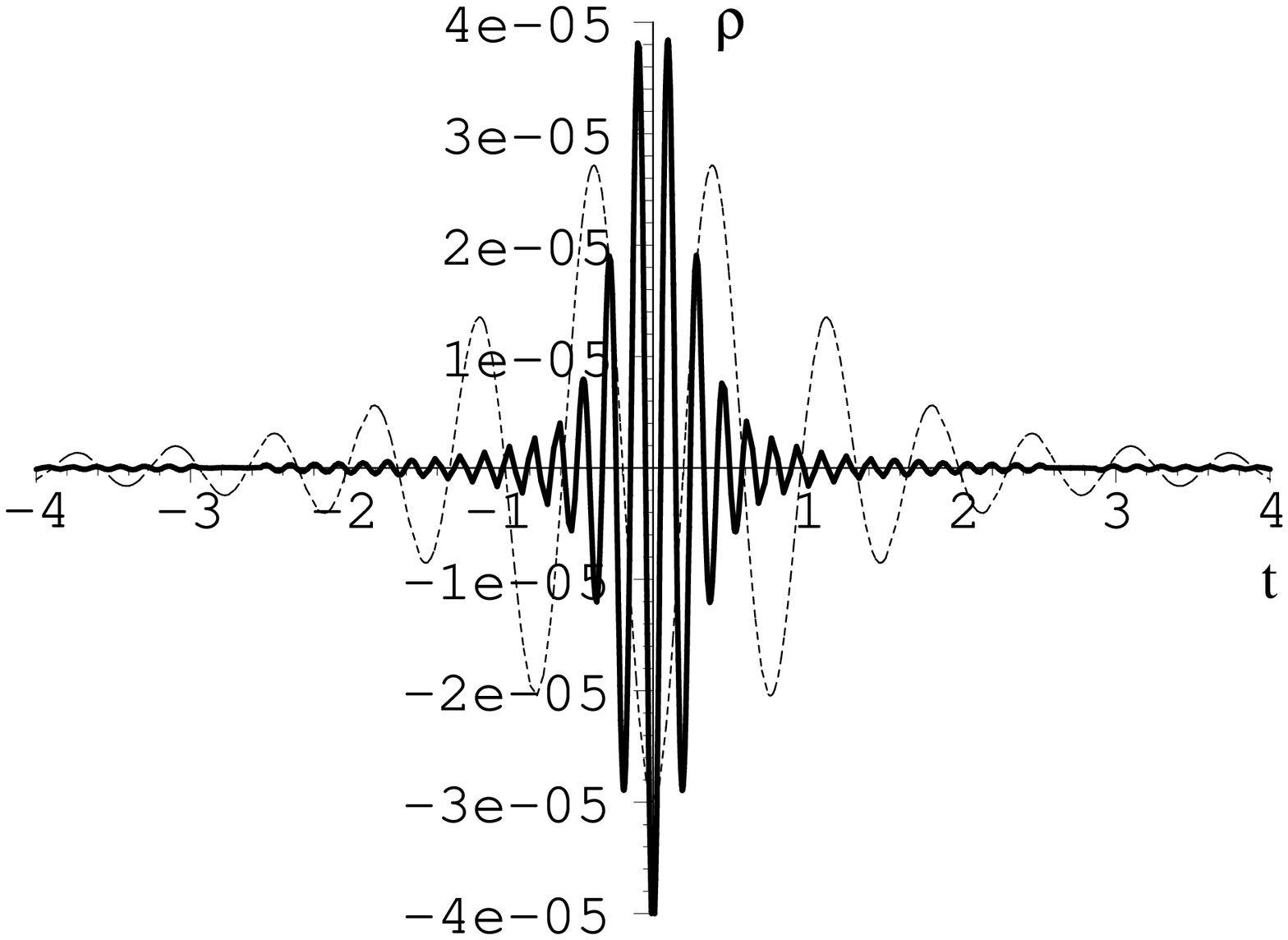}}}}     
\uput[270](1.5,0.5){(a)}\uput[270](4.7,0.5){(b)}
\end{pspicture}
\end{center}
\caption{Plots of the null-contracted stress-energy along (a)~the null
geodesic
$\lambda\mapsto\langle\rho(\lambda,0,0,\lambda)\rangle_{\omega_\alpha}$
for $\alpha=0.2$ (dotted) and $\alpha=0.005$ (solid); (b)~the timelike curve
$t\mapsto \langle\rho(t,0,0,0)\rangle_{\omega_\alpha}$ for $\alpha=0.2$
(dotted) and $\alpha=0.05$ (solid).
In all cases, $\Lambda_0=1$, $\sigma=3.75$.}
\label{fig:worldlines}
\end{figure}

Further insight may be gained from Fig.~\ref{fig:worldlines}, in which we
plot
$\langle :T_{ab}\ell^a\ell^b:\rangle_{\omega_\alpha}$ along (a)~the null
line
$(\lambda,0,0,\lambda)$ and (b)~the timelike line $(t,0,0,0)$.
Along the null line, the effect of decreasing $\alpha$ is essentially
to modify the amplitude of the curve while leaving its shape
substantially unaltered. For a sampling function $f$ supported within
the central trough, it is clear that $\langle
\rho(f)\rangle_{\omega_\alpha}\to-\infty$ as $\alpha\to 0$, in
accordance with Theorem~\ref{thm:main}. Along the timelike curve,
however, decreasing $\alpha$ increases both the amplitude and
frequency of the oscillations. Averaged against a fixed sampling
function, one might expect that the rapid oscillations would tend to
cancel, so that the averages $\int f(t) \langle
\rho(t,0,0,0)\rangle_{\omega_\alpha}$ could be bounded below. This is
borne out by the results of Sec.~\ref{sect:nontriv} below. 

The behavior of the energy density in the vicinity 
of our chosen null geodesic, e.g., as exhibited in Fig.~\ref{fig:density_plots}, 
is almost exactly analogous to that 
found in the analysis of Ref.~\cite{FHR} for spatially averaged QIs, 
in the following sense. There, it was shown that the 
sampled energy density could be unboundedly negative 
in a spatially compact region on a $t$=const surface, 
in four-dimensional Minkowski spacetime. However, for the ordinary 
worldline QIs to hold, the energy density must fluctuate wildly 
as one moves off the $t$=const surface. In the present paper, we find 
a similar result for null rays. 
For example, the central trough in Fig.~\ref{fig:worldlines}(a) is a 
compactly supported region of the null geodesic, analogous to the 
compactly supported spatially sampled region considered in Ref.~\cite{FHR}, 
where the energy density can be made unboundedly negative. 
(Of course, in the null case we are considering a one-dimensional 
average along a line, as opposed to a three-dimensional spatial average).
As we move off the null geodesic, as shown in
Fig.~\ref{fig:worldlines}(b) and Fig.~\ref{fig:density_plots}, the energy
density oscillates rapidly in sign, which must happen if the worldline
QIs are to be satisfied. We speculate that a rotation of the plots in
Fig.~\ref{fig:density_plots}, which makes the white and dark lines
horizontal, would yield a representative picture of the behavior 
in the spatial case. The null and spatial cases seem intuitively to be
very similar.

\subsection{Convergence to the vacuum state} \label{sect:cvgnce}

We have seen that the massless scalar field in four-dimensional
Minkowski spacetime does not satisfy nontrivial null worldline quantum
inequalities. As described above, this was shown by
considering a sequence of vacuum-plus-two-particle states in which the
three-momenta of excited modes become more and more parallel to the
spatial part $\lb$ of the null vector $\ell^a$ as we take the momentum
cut-off to infinity. A perhaps puzzling feature of our sequence is
that it converges in Fock space to the vacuum vector. How, then, can
the energy density diverge?

The answer to this question resides in the fact that the averaged energy
density is an unbounded quadratic form, so the convergence of
a sequence of states in the Hilbert space norm does not imply the
convergence of the corresponding expectation values.
 As a more familiar example,
consider the quantum mechanics of a single harmonic oscillator with
angular frequency $\omega$. Let
\begin{equation}
\phi_n = \ket{0} + n^{-1/4}\ket{n}\,,
\end{equation}
where $n=1,2,3,\ldots$ and $\ket{n}$ is a normalised eigenstate of
energy $\hbar\omega(n+\frac{1}{2})$. Noting that $\|\phi_n\|^2 =
1+n^{-1/2}$, the expected energy is
\begin{equation}
\langle H\rangle_{\phi_n} =
\frac{1}{2}\hbar\omega\frac{1+n^{-1/2}(2n+1)}{1+n^{-1/2}}
=\left[n^{1/2}-\frac{1}{2}+O(n^{-1/2})\right]\hbar\omega
\end{equation}
and therefore diverges as $n\to\infty$, while $\phi_n$ manifestly converges
to the ground state $\ket{0}$, because $\|\phi_n-\ket{0}\| = n^{-1/4}\to 0$.

\subsection{Consistency with the ANEC}

Although, as we have seen, null worldline QNEIs do not exist, there is
nonetheless a nontrivial restriction on the null-contracted stress
energy, namely the averaged null energy condition (ANEC)
\begin{equation}
\int d\lambda\,\langle
:T_{ab}:\ell^a\ell^b\rangle_\omega(\gamma(\lambda)) \ge 0
\end{equation}
established by Klinkhammer~\cite{K} at least for a dense set of states
in the Fock space of the Minkowski vacuum, and by Wald and Yurtsever~\cite{WY91} for a
large subclass of Hadamard states~\footnote{Had QNEIs existed,
one could have derived the ANEC as a consequence, just as the AWEC may be
derived from the QWEIs (see Ref.~\cite{FR95}).  However, the reverse
implication is not valid, so there is no contradiction between
nonexistence of QNEIs and the validity of the ANEC.}. As a consistency
check, we now show explicitly that each state $\omega_\alpha$ obeys
the ANEC, regarded as the requirement that
\begin{equation}
\liminf_{\lambda_0\to+\infty} \frac{1}{f(0)} \int d\lambda\,
f(\lambda/\lambda_0) \langle
:T_{ab}:\ell^a\ell^b\rangle_{\omega_\alpha}(\gamma(\lambda)) \ge 0
\label{eq:ANEC}
\end{equation}
for any $f$ satisfying the hypotheses stated above
Theorem~\ref{thm:main}, and $f(0)\not=0$. Now $\lambda\mapsto f(\lambda/\lambda_0)$ has
Fourier transform $v\mapsto \lambda_0 \widehat{f}(\lambda_0 v)$, which
converges to $2\pi f(0)\delta(v)$ as $\lambda_0\to\infty$. 
Replacing $\widehat{f}$ by this distribution in Eqs.~(\ref{eq:di0}) and~(\ref{eq:di2}), 
we see that, in the limit $\lambda_0\to\infty$,
\begin{equation}
\varphi(v,v')\to 2\pi f(0) \Re \int_0^v du\int_0^{v'} du'\, uu' B(u,u') \delta(u+u') = 0\,,
\end{equation}
while
\begin{eqnarray}
\psi(v,v') &\to&  2\pi f(0) \Re \int_0^v du\int_0^{v'} du'\, uu' C(u,u') \delta(u-u') \nonumber\\
&=& 2\pi f(0)\int_0^{\min\{v,v'\}}
du\,u^2 C(u,u) \nonumber \\
&\ge& 0\,,
\end{eqnarray}
from which Eq.~(\ref{eq:ANEC}) follows. This may also be confirmed by a
more careful analysis.

\begin{figure}
\begin{center}
\resizebox*{5 in}{!}{\includegraphics{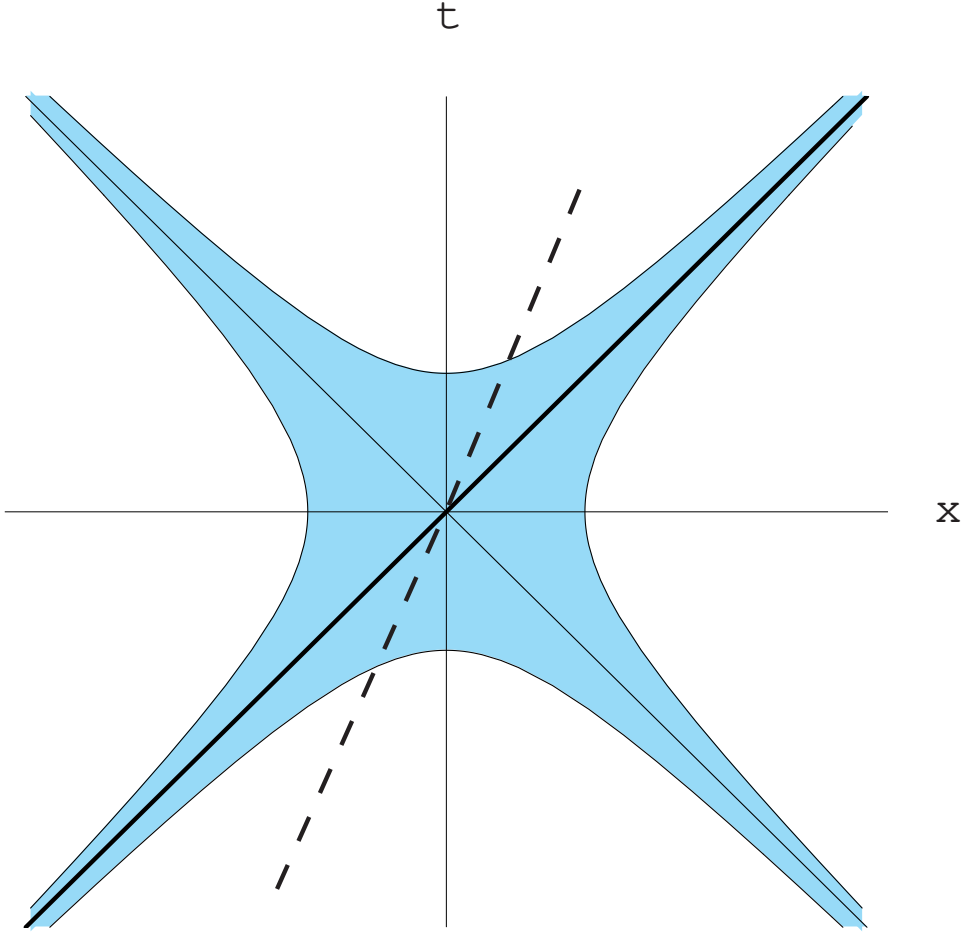}}
\end{center}
\caption{An example to illustrate the distinction between the AWEC and
the ANEC. The shaded region consists of spacetime points $x$ with  $h(x^a x_a)<0$. 
(Only one spatial dimension is shown). The tensor $T_{ab}$ obeys the AWEC
along any timelike geodesic (e.g. the dotted line) but fails to obey the ANEC on any 
null geodesic through the origin (e.g., the solid line).}
\label{fig:counterexample}
\end{figure}

At this point, we take the opportunity to clarify an issue relating to the
derivation of the ANEC given in Ref.~\cite{FR95}, in which it was suggested that
(in Minkowski space) the ANEC could be derived by first taking
the infinite sampling time limit of the QWEI to obtain the AWEC,
and then taking the null limit to conclude that the ANEC holds. However, the following
example shows that the second step cannot be accomplished without further assumptions:
define a function $h(z)$ such that $h(z)$ equals $+1$ for $|z|>2$, and $-1$
for $|z|<1$
with $h(z)$ otherwise smooth and bounded between $\pm 1$. Setting
$t^a=(1,\Ob)$,
\begin{equation}
T_{ab}(x) = t_a t_b h(x^c x_c)
\end{equation}
is a symmetric tensor which satisfies the AWEC along any timelike
geodesic, but fails to satisfy the ANEC along any null generator of the
lightcone at the origin. See Fig.~\ref{fig:counterexample}.  Although
it is not clear to us whether a {\em conserved} tensor field could
display this behaviour, our example shows --- even in Minkowski space --- that
the ANEC cannot be obtained from the AWEC without more assumptions than used
in Ref.~\cite{FR95}.

\section{Timelike worldline QNEIs}
\label{sect:nontriv}

Theorem~\ref{thm:main} may
appear to suggest that null-contracted stress energy tensors are not
subject to any constraints in quantum field theory. This is by no means
the case. Let $(M,\gb)$ be any globally hyperbolic spacetime and
$\ell^a$ a smooth null vector field defined on a tubular neighbourhood
of a smooth timelike curve $\gamma$, parametrized by its proper time
$\tau$. Let $\omega_0$ be any Hadamard state of the Klein--Gordon field
$\Phi$ of mass $m\ge 0$.

\begin{Thm} For any smooth, real-valued, compactly supported function
$g$, the inequality~\footnote{Note that Ref.~\cite{AGWQI} used a
different convention for the Fourier transform in which
Eq.~(\ref{eq:ntm}) would involve $\widehat{F}(-\alpha,\alpha)$ rather
than $\widehat{F}(\alpha,-\alpha)$.}
\begin{equation}
\int d\tau\,\langle :T_{ab}:\ell^a\ell^b\rangle_\omega(\gamma(\tau))
g(\tau)^2
\ge - \int_0^\infty \frac{d\alpha}{\pi} \widehat{F}(\alpha,-\alpha)
\label{eq:ntm}
\end{equation}
holds for all Hadamard states $\omega$ of the Klein--Gordon field of
mass $m$, where normal ordering is performed relative to the state
$\omega_0$ and
\begin{equation}
F(\tau,\tau') = g(\tau)g(\tau')
\langle (\ell^a\nabla_a\Phi)(\gamma(\tau))
(\ell^{b'}\nabla_{b'}\Phi)(\gamma(\tau'))
\rangle_{\omega_0}\,.
\label{eq:Fdef}
\end{equation}
\label{thm:nontriv}
\end{Thm}
{\noindent\bf Remark:} Because the differentiated two-point function
\begin{equation}
H(x,x') = \langle (\ell^a\nabla_a\Phi)(x)
(\ell^{b'}\nabla_{b'}\Phi)(x')
\rangle_{\omega_0}
\end{equation}
is a distribution it is not clear {\em a priori} that one can restrict it
to the curve $\gamma$ as we have done in Eq.~(\ref{eq:Fdef})~\footnote{
As an example, consider the distribution $u(x,y)=\delta(x)$, which has
a sensible restriction $u(x,y_0)=\delta(x)$ to lines of the form $y=y_0$, but no
well-defined restriction to the line $x=0$.}. Techniques
drawn from microlocal analysis provide sufficient conditions for this
to be accomplished, which are satisfied for timelike $\gamma$ owing to
the singularity properties of Hadamard states --- see Ref.~\cite{AGWQI} for
more details on this point. However, the sufficient conditions would not
be satisfied if $\gamma$ was null, which explains why one cannot derive
null worldline QNEIs using the arguments of Ref.~\cite{AGWQI} (although
this does not in itself demonstrate the nonexistence of such bounds).\\
{\noindent\em Proof:} The argument is identical to that used for the
QWEI derived in Ref.~\cite{AGWQI}, in which the averaged quantity was
$\langle :T_{ab}:v^a v^b\rangle_\omega$ (where $v^a$ is the tangent
vector to $\gamma$). We refer to Ref.~\cite{AGWQI} for the details. $\square$

The above bound can be made more quantitative if we return to
four-dimensional Minkowski space, with $\omega_0$ chosen to be the
Poincar\'e invariant vacuum, $\gamma$ chosen to be the worldline of
an inertial observer with four-velocity $v^a$ and with $\ell^a$
some constant null vector field.
By Poincar\'e invariance we may write
$\gamma(\tau)=(\tau,0,0,0)$ without loss of generality; in this frame of
reference, we write $\ell^a=(\ell^0,\lb)$, with $\ell^0 = v^a \ell_a$. We
have
\begin{equation}
H(x,x') = \int\frac{d^3\kb}{(2\pi)^3} \,\frac{(\ell^a k_a)^2}{2\omega}
e^{-ik_a(x^a-{x'}^a)}\,,
\end{equation}
from which it follows that
\begin{equation}
F(\tau,\tau') = g(\tau)g(\tau') \int\frac{d^3\kb}{(2\pi)^3} \,\frac{(\ell^a
k_a)^2}{2\omega}
e^{-i\omega(\tau-\tau')}
\end{equation}
and
\begin{eqnarray}
\widehat{F}(\alpha,-\alpha) &=&
\int d\tau\,d\tau' \int\frac{d^3\kb}{(2\pi)^3} \,\frac{(\ell^a k_a)^2}{2\omega}
e^{-i(\omega+\alpha)(\tau-\tau')}g(\tau)g(\tau')\nonumber\\
&=& \int\frac{d^3\kb}{(2\pi)^3} \,\frac{(\ell^a k_a)^2}{2\omega}
\widehat{g}(\omega+\alpha)\widehat{g}(-\omega-\alpha)\nonumber\\
&=&
\int\frac{d^3\kb}{(2\pi)^3} \,\frac{(\ell^a k_a)^2}{2\omega}
\left|\widehat{g}(\alpha+\omega)\right|^2\,,
\end{eqnarray}
where we have used the fact that $\widehat{g}(-u)=\overline{\widehat{g}(u)}$ 
since $g$ is real. Introducing
polar coordinates about $\lb$ and changing
variables from $k$ to $\omega=\sqrt{k^2+m^2}$, we have
\begin{eqnarray}
\widehat{F}(\alpha,-\alpha)
&=& \frac{(\ell^0)^2}{8\pi^2}
\int_0^\infty dk\,\frac{k^2}{\omega}
\left|\widehat{g}(\alpha+\omega)\right|^2 \int_{-1}^1 d(\cos\theta)\,
(\omega-k\cos\theta)^2\nonumber\\
&=& 
\frac{(\ell^0)^2}{12\pi^2}
\int_0^\infty dk\,\frac{k^2}{\omega}
\left|\widehat{g}(\alpha+\omega)\right|^2 (3\omega^2+k^2)\nonumber\\
&=&
\frac{(\ell^0)^2}{12\pi^2}
\int_m^\infty d\omega\, (\omega^2-m^2)^{1/2}(4\omega^2-m^2)
\left|\widehat{g}(\alpha+\omega)\right|^2\,.
\end{eqnarray}
The right-hand side of the bound~(\ref{eq:ntm}) is thus
\begin{equation}
 - \int_0^\infty \frac{d\alpha}{\pi} \widehat{F}(\alpha,-\alpha)
= -\frac{(v^a\ell_a)^2}{12\pi^3}\int_m^\infty du\,
\left|\widehat{g}(u)\right|^2 \int_m^u  d\omega\,
(\omega^2-m^2)^{1/2}(4\omega^2-m^2) \,,
\end{equation}
so the quantum inequality is
\begin{equation}
\int d\tau\,\langle :T_{ab}:\ell^a\ell^b\rangle_\omega(\gamma(\tau))
g(\tau)^2
\ge -\frac{(v^a\ell_a)^2}{12\pi^3}\int_m^\infty du\,
\left|\widehat{g}(u)\right|^2 u (u^2-m^2)^{3/2}\,.
\end{equation}
In the massless case, we have the simpler expression
\begin{eqnarray}
\int d\tau\,\langle :T_{ab}:\ell^a\ell^b\rangle_\omega(\gamma(\tau))
g(\tau)^2
&\ge & -\frac{(v^a\ell_a)^2}{12\pi^3}\int_0^\infty du\,
u^4\left|\widehat{g}(u)\right|^2 \nonumber\\
&=&-\frac{(v^a\ell_a)^2}{12\pi^2} \int_{-\infty}^\infty d\tau\,g''(\tau)^2\,,
\end{eqnarray}
where we have used Parseval's theorem and the fact that
$|\widehat{g}(u)|$ is even. This takes the same
form as the corresponding QWEI derived in~\cite{FewsterEveson} which reads
\begin{equation}
\int d\tau\, \langle:T_{ab} v^a v^b:\rangle_\omega(\gamma(\tau)) g(\tau)^2 
\ge -\frac{1}{16\pi^2}\int_{-\infty}^\infty d\tau\,g''(\tau)^2 
\end{equation}
in our present notation. For nonzero mass, the two bounds differ by more
than
just an overall factor.

To give a specific example, suppose
that
\begin{equation}
g(\tau) = \left(2\pi\tau_0^2\right)^{-1/4} e^{-\frac{1}{4}(\tau/\tau_0)^2}\,,
\end{equation}
so that $g(\tau)^2$ is a normalised Gaussian with mean zero and variance
$\tau_0>0$. For massless fields, we obtain
\begin{equation}
\int d\tau\,\langle :T_{ab}:\ell^a\ell^b\rangle_\omega(\gamma(\tau))
g(\tau)^2 \ge
-\frac{(v^a\ell_a)^2}{64\pi^2\tau_0^4}\,,
\end{equation}
at least for Hadamard states for which the integral on the left-hand side
converges absolutely~\footnote{This qualification is required in order to
extend the result of Thm.~\ref{thm:nontriv} to noncompactly supported
$g$. To obtain a statement valid for all Hadamard states, one could
alternatively
replace the left-hand side by
${\rm lim\,inf}_{\lambda\to +\infty} \int d\tau\,
\langle :T_{ab}:\ell^a\ell^b\rangle_\omega(\gamma(\tau))
\left[g(\tau) \varphi(\tau/\lambda)\right]^2$
where $\varphi(s)$ is smooth, equal to $1$ for $|s|<1$, vanishing
for $|s|>2$ and monotone decreasing as $|s|$ increases.}.
We note that both sides of this expression scale by a factor of $\sigma^2$
under $\ell^a\mapsto \sigma\ell^a$.

\section{Conclusion}

We have shown, by explicitly constructing 
a counterexample, that quantum inequalities along null
geodesics do not exist in four-dimensional Minkowski
spacetime, for the massless minimally coupled scalar field.
By contrast, it was shown in Ref.~\cite{FR95} that
such bounds do exist in two-dimensional flat
spacetime. The quantum states used in our analysis
are superpositions of the vacuum and multimode two-particle states 
in which the excited modes are those whose three-momenta
lie in a cone centered around our chosen null vector.
We considered the limit of a sequence of such states in which the
three-momenta become arbitrarily large while the radius of the cone
shrinks to zero. Because the dominant contribution arises from
modes with large three-momenta, we expect this result to hold for
massive fields as well.

An interesting feature of our example is that
the sampled energy density along the null geodesic
becomes unbounded from below while the sequence of quantum states converges
to the vacuum state. We demonstrated how such behavior is possible
by considering an analogous example involving the simple harmonic
oscillator in ordinary quantum mechanics. It was also shown that,
as expected, the renormalized stress energy in our class of states
satisfies the ANEC.

As we have learned from Verch (private communication, based on
a remark of Buchholz) our result may be understood as a consequence
of the fact that, in any algebraic quantum field theory in
Minkowski space of dimension $d>2$ obeying a minimal set of reasonable
conditions~\footnote{In more detail, we consider theories described by
von Neumann algebras $R(\OO)$, indexed by open sets $\OO$ of Minkowski
space and consisting of bounded operators on a Hilbert space $\HH$. The
assignment $\OO\mapsto R(\OO)$ is assumed to be {\em isotonous} (i.e.,
$\OO_1\subset\OO_2$ implies that $R(\OO_1)$ is a subalgebra of $R(\OO_2)$) and
{\em local} (i.e., if $\OO_1$ and $\OO_2$ are spacelike separated then
any element of $R(\OO_1)$ commutes with every element of $R(\OO_2)$). 
In addition, the group of translations is implemented on $\HH$ by unitary
operators $U(a)$ such that $U(a)R(\OO)U(a)^{-1}=R(\OO+a)$ and which 
obey the {\em spectral condition}: $U(a)=\int_{\RR^d} dE(p)\, e^{-ia\cdot
p}$ where the spectral measure $dE(\cdot)$ is supported in the forward
light cone $\{p: p\cdot p \ge 0,~p^0\ge 0\}$. It is also assumed
that there is a unique translationally invariant vacuum state
$\Omega\in\HH$ and that $\Omega$ is {\em cyclic} (i.e., $R(\OO)\Omega$ is
dense in $\HH$ for each $\OO$).}
there are no nontrivial observables localised on any
bounded null line segment~\footnote{Under the assumptions in
the previous endnote, let $K$ be a compact subset of Minkowski
space such that $(p-q)\cdot (p-q)\le 0$ for all $p^a,q^a\in K$. 
Suppose that there exist vectors $w_1^a$ and $w_2^a$ such that
$w_1\cdot w_1=w_2\cdot w_2 =-1$, $(w_1-w_2)\cdot (w_1-w_2)\ge 0$ and
with the property that, for all $p^a,q^a\in K$, $|(p-q)\cdot w_i|\le
\sqrt{-(p-q)\cdot(p-q)}$ ($i=1,2)$. Then a theorem of
Woronowicz~\cite{Woron68} entails that the algebra $\bigcap_{\OO\supset
K} R(\OO)$ of observables localised on $K$ consists only of scalar
multiples of the identity. In the case of a bounded null line segment
$K=\{\lambda\ell^a:\lambda\in [\alpha,\beta]\}$ where $\ell^a$ is
null, it is readily verified that $w_1^a=\ell^a+s^a$,
$w_2^a=-\ell^a+s^a$ satisfy these conditions for any spacelike vector
$s^a$ with $s\cdot s=-1$ and $\ell\cdot s=0$. Such vectors $s^a$ exist
in dimensions $d>2$, thus justifying the claim made in the text.}.
Given further reasonable
conditions (cf.~\cite{Verch_ANEC}) this could provide a general
argument for the nonexistence of null worldline QNEIs even
for interacting field theories.

Our results imply that it is not possible to prove
a singularity theorem, such as Penrose's theorem \cite{P},
by using a null worldline QNEI instead of, say, the NEC or the ANEC.
Although our results have been proven only for flat spacetime,
we have no reason to believe that a null worldline QNEI is any more likely
to exist in curved spacetime.

Singularity theorems such as Penrose's theorem involve the focussing
of null geodesics which generate the boundary of the future
of a closed trapped surface. The latter initiates convergence of
a bundle of null rays, and then some bound on the stress-tensor,
such as the NEC, is required to maintain the focussing.
Various sufficient conditions for focussing have
been suggested in the literature \cite{T,C-E,Galloway,B87,TR-88,Y95}. 
However, it should be pointed out that
{\it no} bound on the stress
energy tensor which is strong enough to ensure
sufficient focussing to guarantee
the existence of conjugate points on
half-complete null geodesics (as required in the Penrose theorem)
could hold everywhere in an evaporating
black hole spacetime. Such a bound would be inconsistent with
the existence of Hawking evaporation \cite{H75}. (See Sec.~IV of Ref.~\cite{FR96_EBH}
for a more detailed discussion of this point.) To obtain a singularity,
however, it is only necessary that the required focussing condition hold
for {\it at least one} trapped surface. It is somewhat difficult
to see how one would prove that such a trapped surface would always exist.
As suggested in
Refs.~\cite{FR96_EBH,Yu}, in regions of evaporating
black hole spacetimes where the ANEC is violated,
it may be possible to get a (more limited) QI-type bound that
measures the degree of ANEC violation and which
is also invariant under rescaling of the affine parameter.
Alternatively, one might argue that on dimensional grounds,
the curvature which promotes focussing scales as ${l_c}^{-2}$,
where $l_c$ is the local proper radius of curvature,
while the energy densities produced by quantum fields typically
scale only like ${l_c}^{-4}$. If this line of reasoning is
correct, then one might expect the breakdown of the energy conditions
to only affect the validity of the singularity theorems when
$l_c=l_{Planck}$ \cite{H75}.

In this paper, we also showed that---in general globally hyperbolic
spacetimes---averages of null-contracted stress-energy
of massive and massless fields along {\em timelike} curves are constrained by 
quantum inequalities. Large negative energy
densities concentrated along a null geodesic must therefore be compensated
by large positive energy densities on neighbouring null geodesics.
This is reminiscent of the transverse smearing employed by Flanagan
and Wald \cite{FW} in their study of the ANEC in semiclassical quantum
gravity. Such transversely smeared observables also evade the
Buchholz--Verch argument mentioned above.
Whether physically interesting global results, such as singularity
theorems, can be proved using inequalities such as~(\ref{eq:ntm}) is an open
question, which is currently under investigation.

\begin{acknowledgments}
The authors thank Rainer Verch for
raising the issue mentioned in the conclusion and Detlev Buchholz
for further discussions on this point and help in locating Ref.~\cite{Woron68}.
Thanks are also due to Klaus 
Fredenhagen for useful conversations and Mitch Pfenning for help in preparing some of
the figures. This research was partly conducted at the Erwin Schr\"odinger Institute in
Vienna during the programme on Quantum Field Theory in Curved Spacetime;
we are grateful to the ESI for support and hospitality. 
TAR is also grateful to the Mathematical Physics group at
the University of York for hospitality during the early phases of the
work. This research was supported in part by EPSRC grant~GR/R25019/01 
to the University of York (CJF) and NSF grant No.~Phy-9988464 (TAR).
\end{acknowledgments}

\end{document}